\documentclass[fleqn,usenatbib]{mnras}
\usepackage{longtable}
\usepackage{newtxtext,newtxmath}

\usepackage[T1]{fontenc}

\DeclareRobustCommand{\VAN}[3]{#2}
\let\VANthebibliography\thebibliography
\def\thebibliography{\DeclareRobustCommand{\VAN}[3]{##3}\VANthebibliography}

\usepackage{graphicx}	
\graphicspath{{figures/}}
\usepackage{pdflscape}
\usepackage[dvipsnames]{xcolor}

\title[XTE J1550-564 and GRO J1655-40]{Testing General Relativity using Quasi-Periodic Oscillations from X-ray Black Holes: XTE J1550-564 and GRO J1655-40}

\author[Rink, Caiazzo \& Heyl]{
Katherine Rink\thanks{Current address: UMass Dartmouth; email: krink@umassd.edu}$^1$
Ilaria Caiazzo\thanks{email: ilariac@caltech.edu; Sherman Fairchild Fellow}$^2$, Jeremy Heyl\thanks{email: heyl@phas.ubc.ca}$^1$,
\\
$^{1}$Department of Physics and Astronomy, University of British Columbia, Vancouver, BC V6T 1Z1, Canada\\
$^{2}$TAPIR, Walter Burke Institute for Theoretical Physics, Mail Code 350-17, Caltech, Pasadena, CA 91125, USA\\
}

\date{Accepted XXX. Received YYY; in original form ZZZ}

\pubyear{2021}

\begin{document}
\label{firstpage}
\pagerange{\pageref{firstpage}--\pageref{lastpage}}
\maketitle

\begin{abstract}

We use the Relativistic Precession Model (RPM) \citep{RPM} and quasi-periodic oscillation (QPO) observations from the \textit{Rossi X-ray Timing Explorer} to derive constraints on the properties of the black holes that power these sources and to test General Relativity (GR) in the strong field regime. We extend the techniques outlined by \citet{mottaa,mottab} to use pairs of simultaneously measured QPOs, rather than triplets, and extend the underlying spacetime metric to constrain potential deviations from the predictions of GR for astrophysical black holes.  To do this, we modify the RPM model to a Kerr-Newman-deSitter spacetime and model changes in the radial, ecliptic, and vertical frequencies.  We compare our models with X-ray data of XTE J1550-564 and GRO J1655-40 using robust statistical techniques to constrain the parameters of the black holes and the deviations from GR.  For both sources we constrain particular deviations from GR to be less than one part per thousand.
\end{abstract}
 

\section{Introduction}


The X-ray flux from accreting black holes presents interesting nearly periodic fluctuations, called quasi-periodic oscillations or QPOs, that are usually found upon inspection of the power density spectrum (PDS) \citep{1991ApJ...374..741M,1997ApJ...489..272T,2005Ap&SS.300..149V,2006csxs.book...39V}. Such fluctuations are detected as narrow peaks in the PDS, and often appear at three distinct frequencies. When three frequencies are observed simultaneously, they usually consist of two higher frequencies ($\sim$100 Hz) and one lower frequency (ranging from a few mHz to $\sim$ 30 Hz) \citep{1710}.


The true nature of QPOs is still unknown, but the frequencies detected are similar to the orbital and epicyclic timescales of the inner accretion disk of a Kerr black hole, hinting to the possible origin of these phenomena. Several models have been proposed to explain QPOs, and they have mostly focused on either the low-frequency or the high-frequency oscillations. For the low-frequency QPOs, the most commonly invoked mechanisms are either some instabilities in the accretion flow or geometric oscillations \citep{1993ApJ...417..671C,1999A&A...349.1003T,2010MNRAS.404..738C}, and in particular Lense-Thirring precession \citep{1998ApJ...492L..59S,1999ApJ...524L..63S,2001ApJ...559L..25W,2006ApJ...642..420S,2009MNRAS.397L.101I}. The latter consists of a nodal precession of orbits inclined to the equatorial plane caused by a spinning compact object dragging the surrounding spacetime around with it (i.e. the frame dragging effect). The high-frequency oscillations have been observed more rarely in black hole systems, and are more obscure; the proposed models include Doppler modulation of orbiting hotspots in the inner disk, oscillation modes of a pressure-supported torus, nonlinear resonances, gravity and pressure modes in the accretion disk \citep[see][and references therein]{2013IAUS..290...57L,2014SSRv..183...43B}.


One of the models that tries to explain the different, high and low, frequencies together is the Relativistic Precession Model (RPM) \citep{RPM}.
The theory argues that the coronal disk (the innermost part of the accretion disk) is perturbed and tilted out of the equatorial plane. The three QPOs would then appear in the PDS if a particularly hot region in the coronal disk formed due to random perturbations in temperature.  As the node of the orbit precesses, the flux will vary slowly at the nodal precession frequency. The fastest variation would then be at the Kepler frequency, corresponding simply to a hot blob of material moving through its orbit. Finally, the intermediate frequency would correspond to the periastron precession. Given the geometry of elliptical orbits one would expect the material to be compressed most at the periastron, so as this location precesses about the black hole, the flux could vary at this frequency as well. This would result in the observed object's PDS showing peaks in these three frequencies. 

For the X-ray binaries XTE J1550-564 and GRO J1655-40, \citet{mottaa,mottab} published a catalogue of the QPO frequencies observed in these systems. In the GRO~J1655-40 PDS, three frequencies are observed simultaneously. Under the assumption that all three frequencies result from an oscillation at the same radius in the accretion disk, \citet{mottaa} derived the mass and spin of the black hole, and the radius of the oscillation.  For XTE~J1550-564, \citet{mottab} used a pair of simultaneously observed frequencies, along with an estimate of the black hole mass from other observations, to derive constraints on the spin of the black hole and the radius of the oscillation.  In both studies, they used the derived values to calculate the expected frequencies for a range of radii and compared these values with other observations for each source. Here we will take this idea to its natural conclusion.

In both sources, \citet{mottaa,mottab} present about 30 simultaneous measurements of QPOs where a high frequency and a low frequency signal were observed simultaneously.  We use all these measurements to fit for the mass and spin of each black hole as well as other parameters that we expect to vanish for astrophysical black holes. In our calculations, we assume that the physical properties of the black holes do not change from measurement to measurement, that the RPM picture is the correct explanation for the QPOs observed in these systems, and, when two frequencies are observed simultaneously, that they come from the same region of the accretion disk and we can successfully match them to a particular RPM frequency.  Furthermore, we use a robust fitting technique that is insensitive to outliers while equivalent to a $\chi^2$-test in the limit of normally distributed data.  In this way, we can test the validity of the model and identify observations for which the model may not fit or the frequencies were matched incorrectly.

When one uses a triplet of coincident frequencies as in \citet{mottaa}, one can only determine three values: the mass and spin of the black hole and the radius at which the oscillation occurs.  In principle, one might not be able to find values that are consistent with a triplet, so one could compare the values predicted by the model with separate observations; both of these provide tests of the RPM picture and also of the spacetime within the model.  However, the conclusions drawn rest on the single observation. Here we use the entire suite of observations to constrain the parameters of the black hole, constrain deviations from the expectations of GR, determine whether the observations are well fit by the RPM, and if not, which observations do not fit the RPM.

To look at deviations from GR, we employ a black hole spacetime calculated within the context of GR that retains all of the symmetries of the Kerr spacetime, so that the necessary frequencies are straightforward to calculate in closed form, and yet in general contains additional curvature due to known sources that we expect to be absent in astrophysical black holes; specifically, a significant electromagnetic field or a uniform energy density with an equation of state similar to that of the cosmological constant. One can interpret our results to constrain the energy density  around the black hole with respect to empty space within GR or to constrain the underlying gravitational theory to be close to GR.

Although there is a large literature developing constraints on gravity \citep{2014LRR....17....4W} from experiments  in laboratories \citep{2003ARNPS..53...77A}, the Solar System \citep{2007PhRvD..75l4014C,2007GReGr..39.1381A}, and pulsars in binaries \citep{2003LRR.....6....5S} through the parameterized post-Newtonian (PPN) formalism \citep{1968PhRv..169.1017N,1972ApJ...177..757W}, these results do not translate into understanding gravity in the strong-field limit near a black hole.  Developing a fully consistent alternative theory of relativistic gravity to GR is extremely challenging.  The alternatives sometimes fail to yield even a consistent picture of spacetime, are often only defined for small perturbations about flat spacetimes and rarely result in stable black holes.  One technique to make progress is to alter the black hole metric in a parameterized fashion inspired by the PPN formalism \citep{2016EPJC...76..290G}. 

More recent studies exploring GR alternatives for fitting observed QPO triplets from GRO J1655-40 include using quadratic gravity theories (\cite{1703.01472}), the Einstein-Dilaton-Gauss-Bonnet theory (\cite{1412.3473}), the Bardeen metric and Johannsen-Psaltis metric (\cite{1312.2228}), and quadrupolar metrics such as the Kerr-Q and Hartle-Thorne metrics (\cite{2102.02232}). We take the approach to work within GR but change the sources, which in turn yields to potential deviations in the spacetime that we can hope to measure.

The question of whether GR applies to black holes has very profound implications. Most of our fundamental theories in physics have a range of applicability and break down at some particular scale. For example, the electroweak theory separates into electromagnetism and weak theory at energies lower than $\sim$ 200 GeV, while strong and electroweak theories are thought to unify at very high energies. Since most of the current tests of GR are performed in the weak gravitational fields present in our solar system, the 6 to 7 orders of magnitude higher gravitational potential and $\sim$20 orders of magnitude greater curvature around a black hole provide a more extreme environment to test GR in the strong field regime. Much progress has been made in this sense in the field of gravitational waves. However, the wavelength of gravitational waves is necessarily comparable to the size of the colliding objects, which therefore limits the scope for probing in detail the spacetime surrounding compact objects; therefore, electromagnetic signals provide a crucial complementary window into strong gravity. 

One of the most subtle consequences of GR is the ``no-hair'' theorem, for which black holes can be fully characterized by their mass, angular momentum, and charge \citep{1967PhRv..164.1776I,1968CMaPh...8..245I,1968PhRv..174.1559C,1971PhRvL..26..331C,1971PhRvL..26.1653W}. Because we expect no charge on astrophysical black holes, and we can neglect the mass in the immediate vicinity of the black hole, the spacetime that surrounds a black hole can be nearly exactly described by the Kerr metric. The only way to test this theorem is to probe the spacetime very close to the hole. Fortunately, the X-ray emission of accreting black holes carries information about the inner region of the accretion disk, within a few gravitational radii ($R_g=GM/c^2$) from the hole, encoded in the fast variability of its spectrum.


\section{Geodesics in the Kerr-Newman-de-Sitter Metric}

There exists four primary metrics for describing the theoretical types of black holes. The Schwarzschild metric describes a non-rotating, uncharged black hole; this is the simplest case and is not expected to be common. The Kerr metric is the most widely used for it describes a rotating and uncharged black hole. The Reissner-Nordstr\"om metric describes the least probable scenario, a charged and non-rotating black hole. The metric that is modified in this paper for the derivation of the frequency equations is the Kerr-Newman metric. This metric describes a charged and rotating black hole. 

Black holes can theoretically carry charge, but an accreting astrophysical black hole is immersed in plasma and would discharge quickly, so we expect the observed black holes to have negligible electromagnetic charge. A measurement of the effects of charge on the spacetime surrounding a black hole constrains deviations from GR or, in principle, the presence of additional charged components in the Universe.  Here we add an extra parameter to the Kerr-Newman metric that results from an uniform energy density near the black hole with an equation of state similar to that of the cosmological constant.  These constituents yield the Kerr-Newman-deSitter Metric \citep{1973blho.conf...57C,2009GReGr..41.2873C}:
\begin{equation}
\begin{split}
    ds^2  =\frac{\Delta_r^2}{\rho^2} & \left(dt-\frac{a}{\Xi}\sin^2(\theta)d\phi\right)^2 - \frac{\rho^2}{\Delta_r^2}dr^2 - \frac{\rho^2}{\Delta_\theta} d\theta^2- \\ 
    &  ~~~ \frac{\Delta_\theta \sin^2(\theta)}{\rho^2} \left (adt-\frac{r^2+a^2}{\Xi} d\phi\right)^2
\end{split}
\label{eq:spacetime}
\end{equation}
where
\begin{eqnarray}
\Delta_r^2 &= &(r^2+a^2) (1-H^2 r^2)-2Mr+Q^2 \\
\Delta_\theta &= &1+H^2 a^2 \cos^2(\theta) \\
\Xi &=& 1+H^2 a^2 \\
\rho^2 &=& r^2+a^2 \cos^2(\theta) 
\end{eqnarray}
In the above expressions, $t$ is a timelike coordinate, $r$ is a radial coordinate, $a$ is the spin of the black hole, $\theta$ is the angle above or below the plane of the accretion disk, $\phi$ is the angle around the black hole within the plane of the accretion disk, $Q$ is the charge of the black hole, and finally $H^2$ is the cosmological constant parameter. For simplicity, the $t-$coordinate in Eq.~\ref{eq:spacetime}  corresponds to the proper time measured by a distant observer. We will use the value of $Q^2$ throughout the rest of the paper because we interpret $Q^2$ as a parameter to change the structure of the spacetime rather than to quantify the charge of the black hole.  

Furthermore, if the value of $H^2$ does not vanish, the spacetime as given is not asymptotically flat; therefore, Eq.~\ref{eq:spacetime} should be interpreted as a chart that is valid near the black hole in the region where the QPOs originate, typically about ten gravitational radii.  Beyond this region, Eq.~\ref{eq:spacetime} is to be considered attached to another chart that is asymptotically flat. 

In the RPM picture \citep{RPM}, the observed frequencies of the QPOs correspond to three different frequencies in the orbital motion, believed to be the Kepler ($\Omega_{\phi}$), the nodal precession ($\Omega_{\rm node}$), and the periastron precession ($\Omega_{\rm peri}$) frequencies. The Kepler frequency represents the angular frequency at which the object completes one full orbit about the spin axis of the black hole from the point of view of a distance observer. It is defined as the value of $d\phi / dt$ for a geodesic with a constant value of $r$ and $\theta=\pi/2$ (the equatorial plane).  The nodal precession frequency is the difference between the Kepler frequency and the vertical epicyclic frequency ($\omega_\theta$).  The latter is the frequency of oscillations in the $\theta$-direction about the orbits with constant $r$ in the equatorial plane.  This can be thought of as the precession of the two points at which the tilted orbit intersects the equatorial plane.  The periastron precession frequency is the difference between the Kepler frequency and the radial epicyclic frequency ($\omega_r$) and represents the precession of the point of closest approach, where we'd expect to see peak emission.

The three observed frequencies can be represented as follows
\begin{eqnarray}
    \Omega_{\phi} &=& \frac{u^\phi}{u^t} \\
    \Omega_{\textrm{\scriptsize node}} &=& \Omega_{\phi} - \omega_{\theta} \\
    \Omega_{\textrm{\scriptsize peri}} &=& \Omega_{\phi} - \omega_r\,.
\end{eqnarray}
To derive the three measured QPO frequency equations for this model, their corresponding geodesic equations are needed to find the necessary 4-velocity vector and its perturbations. The geodesic equation is used to derive  $\Omega_{\phi}$, $\omega_r$, and $\omega_{\theta}$.

We begin the derivations for a general stationary, axisymmetric spacetime \cite[see][for a similar treatment]{2020EPJC...80..504G} by finding the Kepler frequency, $\Omega_{\phi}$, using the geodesic equation with respect to radius, $r$. Since $u_{r} = g_{rr}u^r$, and $u^r = \frac{dr}{d\tau}$, we can re-write the geodesic equation using:
\begin{equation}
    \frac{du_r}{d\tau} = \frac{d}{d\tau}g_{rr}u^r = \frac{d}{d\tau} \left (g_{rr}\frac{dr}{d\tau} \right )\,.
\end{equation}
However, since the object's radial velocity is constant over time, we can write the change in this velocity as $\frac{d}{d\tau} \left  ( g_{rr}\frac{dr}{d\tau} \right ) = 0$, simplifying the geodesic equation to:
\begin{equation}
    \frac{1}{2}(g_{\mu\nu,r}u^{\mu}u^{\nu}) = 0\,.
\end{equation}
For the Kepler frequency, we assume that the orbit is at a constant radius in the equatorial plane, so the sum reduces to three terms
\begin{equation}
    \frac{1}{2}(g_{tt,r}u^{t}u^{t} + 2g_{t\phi,r}u^{t}u^{\phi} + g_{\phi\phi,r}u^{\phi}u^{\phi}) = 0\,.
\end{equation}
Because the Kepler frequency is $\Omega_{\phi} = u^\phi/u^t$, dividing the equation by $u^t u^t/2$ yields a quadratic equation for the Kepler frequency
\begin{equation}
    (g_{tt,r} + 2g_{t\phi,r}\Omega_{\phi} + g_{\phi\phi,r}\Omega_{\phi}^2) = 0\,.
    \label{eq:geodesic_circular}
\end{equation}
Using the coefficients from the metric and solving the quadratic equation for the Kepler frequency gives
\begin{equation}
    \Omega_{\phi} = \frac{(Mr - H^2r^4 - Q^2)(1+H^2a^2)}{r^2\sqrt{Mr - H^2r^4-Q^2}-a(Mr - H^2r^4 - Q^2)}\,.
\end{equation}
The complete expression for the Kepler frequency shows where in the spacetime each of three parameters $M$, $Q$ and $H$ play a role. If we take $Q=H=a=0$, we recover Kepler's third law
\begin{equation}
    \Omega_{\phi}^2 = \frac{M}{r^3}\,,
    \label{eq:Mfreq}
\end{equation}
allowing only $Q^2$ to have a finite value while the others vanish yields
\begin{equation}
    \Omega_{\phi}^2 =  -\frac{Q^2}{r^4}\,,
    \label{eq:Qfreq}
\end{equation}
and finally, allowing only $H^2$ to have a finite value yields
\begin{equation}
    \Omega_{\phi}^2 = -H^2.
\label{eq:Hfreq}
\end{equation}
Thus, relative to the mass of the black hole, the charge affects the frequency at smaller radii and $H$ affects the frequency at larger radii.

For the remaining two frequencies, the value of $u^t$ is needed for the circular orbit. We begin by noting that
\begin{equation}
    u \cdot u = u_{\alpha}u^{\beta} = g_{\alpha \beta}u^{\alpha}u^{\beta} = 1 \,.
\end{equation}
Because the nodal and periastron precession frequencies can be thought of as perturbations to the Kepler frequency, a circular orbit is assumed in the derivation of $u^t$. This assumption results in only the $t$ and $\phi$ components playing a role, such that
\begin{equation}
    1 = g_{tt}(u^t)^2 + 2g_{t\phi}u^{\phi}u^t + g_{\phi \phi}(u^{\phi})^2 \,.
    \label{eq:udotu}
\end{equation}
Remembering that $\Omega_{\phi} = u^{\phi} / u^t$, and therefore $u^{\phi} = \Omega_{\phi}u^t$:
\begin{equation}
    1 = g_{tt}(u^t)^2 + 2g_{t\phi}\Omega_{\phi}(u^t)^2 + g_{\phi \phi}(\Omega_{\phi}u^t)^2 \,,
\end{equation}
where the Kepler frequency, $\Omega_{\phi}$, was previously derived. This leaves the above equation with only one unknown, $u^t$, and hence:
\begin{equation}
    u^t = (g_{tt} + 2g_{t\phi}\Omega_{\phi} + g_{\phi \phi}\Omega_{\phi}^2)^{-\frac{1}{2}} \,.
\end{equation}

To find the radial epicyclic frequency, $\omega_r$, we use the conserved quantities for the orbit. The quantity $u_{\phi}$ is the angular momentum ($L$) and $u_t$ is the energy ($E$). With these new definitions, and a corresponding change of indices from lowered to raised on all metric coefficients, we can rewrite Eq.~\ref{eq:udotu} as follows:
\begin{equation}
    1 = g^{tt}E^2 + 2g^{t\phi}EL + g^{\phi \phi}L^2 \,.
\end{equation}
To perturb the orbit,  additional terms are added to account for movement in the radial direction.:
\begin{equation}
    1 + E_2 = g^{tt}E^2 + 2g^{t\phi}EL + g^{\phi \phi}L^2 + g_{rr}(u^r)^2 \,,
\end{equation}
where $u^r$ is the radial component of the four velocity, $E_2$ is the additional energy of the perturbation and the values of $E$ and $L$ are held fixed at the values for a circular orbit. 
We perturb the position about the circular orbit which has $r=r_0$ and use Eq.~\ref{eq:udotu} and the geodesic equation of the circular orbit (Eq.~\ref{eq:geodesic_circular}) to yield the second-order perturbation
\begin{equation}
    E_2 = \frac{\left(r-r_0\right)^2}{2} \left ( g_{,rr}^{tt}E^2 + 2g_{,rr}^{t\phi}EL + g_{,rr}^{\phi \phi}L^2 \right ) + g_{rr}(u^r)^2 \,.
\end{equation}
By dividing the above equation by two, it becomes clear that the $\frac{1}{2}g_{rr}(u^r)^2$ term behaves as kinetic energy (classically of the form $\frac{1}{2}mv^2$). The above equation is therefore a conservation equation, with the remaining terms accounting for the effective potential energy,
\begin{equation}
E_2 = \frac{1}{2} k \left ( r - r_0 \right )^2 + \frac{1}{2} m v^2
\end{equation}
where $v=u^r=dr/d\tau$, $m=g_{rr}$ and
\begin{equation}
    k = \frac{1}{2} \left ( g_{,rr}^{tt}E^2 + 2g_{,rr}^{t\phi}EL + g_{,rr}^{\phi \phi}L^2 \right ) .
\end{equation}
In the frame of the orbiting particle, the oscillation frequency is given by
\begin{equation}
\omega^2 = \frac{k}{m} = 
    \frac{g_{,rr}^{tt}E^2 + 2g_{,rr}^{t\phi}EL + g_{,rr}^{\phi \phi}L^2}{2g_{rr}} \,.
\end{equation}
The next step is to convert everything from proper time to time in the observer's reference frame. After converting reference frames, we are left with our radial solution:
\begin{equation}
    \omega_r^2 = \frac{g_{,rr}^{tt}E^2 + 2g_{,rr}^{t\phi}EL + g_{,rr}^{\phi \phi}L^2}{2g_{rr}(u^t)^2} \,.
\end{equation}
This equation can be simplified further by recognizing that $u^{\phi} = \Omega_{\phi} u^t$ and dividing out the $(u^t)^2$ term in the denominator
\begin{equation}
    \omega_r^2 = \frac{g_{tt,rr} + 2g_{t\phi,rr}\Omega_{\phi} + g_{\phi\phi,rr}\Omega_{\phi}^2}{2g_{rr}} \,.
\end{equation}
Because no assumptions were made in the derivation of $\omega_r$ that would not also apply to $\omega_{\theta}$, the solution for the final frequency $\omega_{\theta}$ can be carried out in the exact same manner but holding $r$ constant and perturbing $\theta$. For brevity, only the final result is included here:
\begin{equation}
   \omega_{\theta}^2 = \frac{g_{tt, \theta \theta} + 2g_{t \phi, \theta \theta}\Omega_{\phi} + g_{\phi \phi, \theta \theta}\Omega_{\phi}^2}{2g_{\theta \theta}}\,.
\end{equation}

\section{Method}

\begin{figure}
    \centering
    \includegraphics[width=\columnwidth]{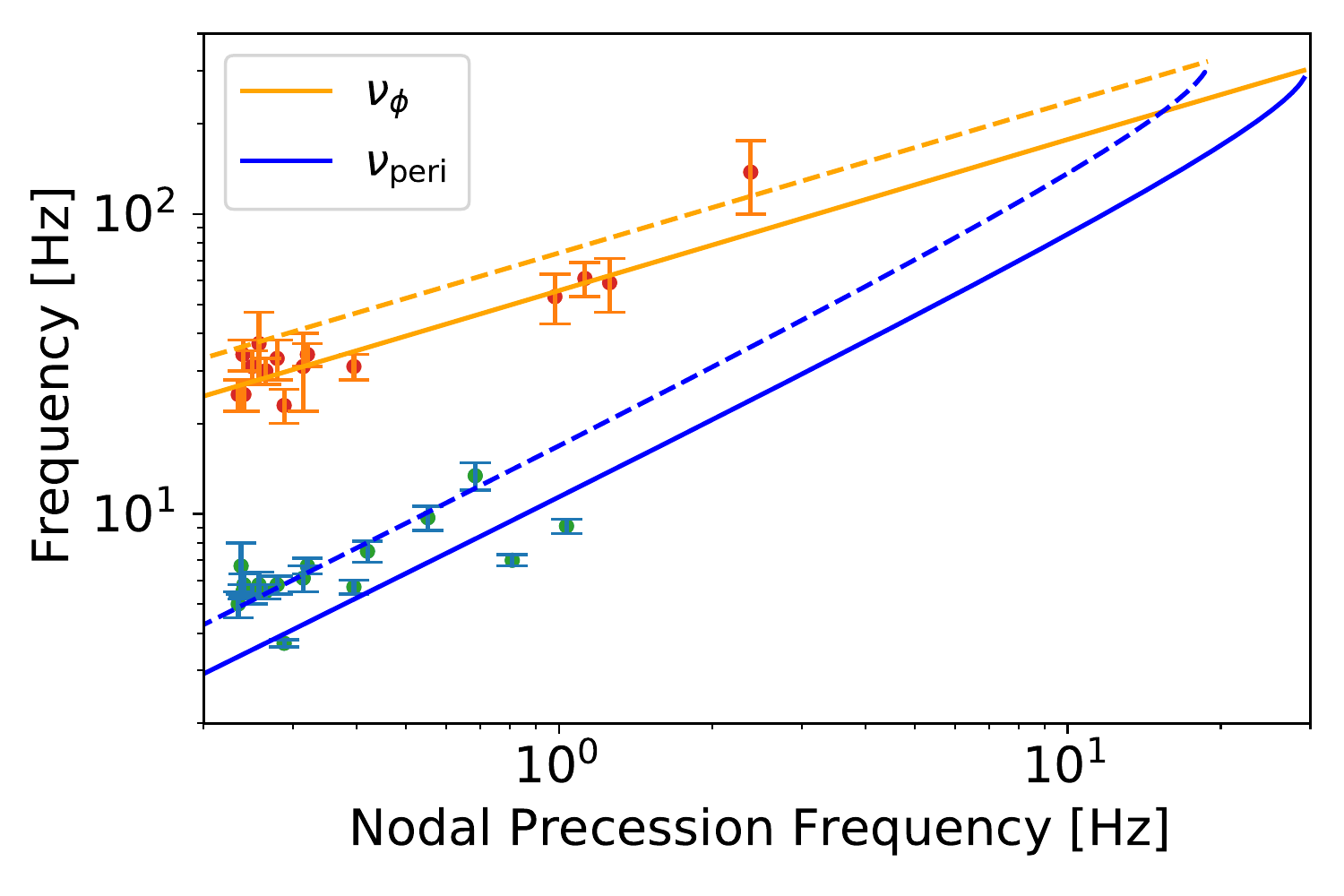}
    \caption{The observed low-frequency QPOs from XTE J1550-564 interpreted to be nodal precession frequencies and the simultaneously observed high-frequency QPOS from \citet{mottab}.  The dashed curves give the values predicted by the RPM for the parameters determined by \citet{mottab}, $M=9.1~\textrm{M}_\odot$ and $a/M=0.34$ ($\chi^2=1456$) .  The solid curves trace the best-fitting model (the smallest value of $\chi^2$, 386) with $M=11.5~\textrm{M}_\odot$ and $a/M=0.49$.  The Kepler frequency and the periastron precession frequency ($\nu_\phi$ and $\nu_{\textrm{\scriptsize peri}}$) are equal to each other at the innermost stable orbit.}
    \label{fig:xte_chi2}
\end{figure}

It is perhaps most straightforward to explain our technique and the nature of the dataset by looking at the frequencies for the source XTE~J1550-564 \citep{mottab}. Fig.~\ref{fig:xte_chi2} depicts for this source the observed frequency pairs, that is, a low-frequency and a high-frequency QPO observed simultaneously.  The low-frequency ones, which are interpreted to be the nodal precession frequency, are depicted along the horizontal axis and the high-frequency ones, which are interpreted as either the Kepler (larger) or periastron-precession (smaller) frequency, along the vertical.  \citet{mottab} also observed many single frequencies which they interpreted as the low-frequency component, not depicted here. They fit the data points, using a prior on the black hole mass of 9.1 $\textrm{M}_{\odot}$, through a five-step process that includes solving for the radius of each QPO given a model of fixed mass and spin and then finding the ensemble of spins consistent with the mass constraint and the observational data. For Fig.~\ref{fig:xte_chi2} we calculate the three angular QPO frequencies as a function of radius for a particular black hole mass and spin.  We recast these calculations as the periastron and Kepler frequencies as a function of the nodal precession frequency and fit the observed pairs of frequencies to the corresponding functions, converting from angular to spatial frequencies.  Unlike \citet{mottab}, we do not use the values of the low-frequency QPO if a high-frequency QPO is not measured simultaneously. 

An examination of Fig.~\ref{fig:xte_chi2} shows that the values of mass and spin determined by \citet{mottab} do not fit all of the data well.  In fact the value of $\chi^2$ is 1456 with 31 degrees of freedom. \citet{mottab} use other observations to fix the mass of the black hole.  This choice forces the values of the Kepler frequency within the model to lie along the upper envelope of the observations.   On the other hand, the periastron precession frequency does trace the bulk of the observations.  The solid curves depict the best-fitting model with a  $\chi^2$ value of 386.  This is a better but still poor fit to the observations.  Although this model traces the bulk of the highest frequency QPOs, it misses the bulk of the middle frequency QPOs.  In this case, it is the QPO observations themselves that create the tension. 

If any of the assumptions underlying the RPM fail, the observed QPOs need not fit the model.  In particular, if some oscillations correspond to frequencies other than the nodal, periastron precession and Kepler frequencies, or if, on occasion, there are perturbations at several radii, then outlying observations will result.  The underlying assumption behind $\chi^2$-fitting of a dataset is that the residuals between the data and the model are normally distributed.  Where the RPM fails for one reason on another, this is not the case.

For this reason, we seek a technique that is robust against the presence of a few outliers in our dataset and yet is efficient in determining the best-fitting model with the fewest observations. To motivate our figure of merit to fit the observed QPO frequencies, we start with the traditional $\chi^2$ estimator,
\begin{equation}
    \chi^2 = \sum_i \left ( X_i \right )^2 = \left ( \sum_j X_j \right )^2 + \sum_i \left ( X_i -  \frac{1}{n} \sum_j X_j  \right )^2
    \label{eq:chi2_traditional}
\end{equation}
where $n$ is the number of measurements and 
\begin{equation}
    X_i = \frac{y_i - f(x_i|\textrm{model parameters})}{\sigma_i}
\end{equation}
is the residual of a particular measurement $y_i$ with respect to the model, inversely weighted by the uncertainty in that measurement.  The first term in Eq.~\ref{eq:chi2_traditional} is the product of the mean value of $X_i$ and the number of measurements, squared.  The second term is the product of the variance in the value $X_i$ and the number of measurements, also squared. To derive a robust figure of merit, we can replace the mean and variance with robust estimators of location and scale. Two such estimators that may be familiar are the median and the median absolution deviation (MAD),
\begin{equation}
    \textrm{MAD} = \textrm{median} \left \{ \left | X_i - \textrm{median} \left( X_i \right ) \right | \right \}.
\end{equation}
Although these two estimators are robust with a breakdown point of fifty percent, they are not efficient, which means that, in the case of normally distributed measurements, one would need many more measurements to obtain the same constraints on the models as one would achieve with the traditional $\chi^2$ estimator. We use two alternative statistics based on pairs of measurements: the Hodges-Lehmann-Sen estimator $H\!L$ \citep{hodges1963,10.2307/2527532} and $Q_n$ \citep{doi:10.1080/01621459.1993.10476408}, which are defined as 
\begin{eqnarray}
H\!L &=&\textrm{median}\left\{\frac{X_{i}+X_{j}}{2} :   i \leq j  \right\}, \\
Q_{n} &=& c_{n}\textrm{~first quartile of~}\left\{\left|X_{i}-X_{j}\right| :  i<j \right\},
\end{eqnarray}
where the value of $c_{n}$ depends on the size of the sample and scales the value of $Q_n$ to be the standard deviation in the case of normally distributed data.  Although both of these estimators appear to require sorting the complete list of pairs and ${\cal O}(n^2 \log n)$ operations, there are efficient techniques\footnote{A python library is available at \url{https://github.com/UBC-Astrophysics/qn_stat}.}, ${\cal O}(n\log n)$, to determine both \citep{doi:10.1080/01621459.1993.10476408,10.1145/1271.319414}.




Because both of these estimators rely on ranks like the median and MAD, they are robust against outliers with breakdown points of 29 and 50 percent, respectively.  A 50 percent breakdown point means that one could change the values of half of the data without changing the estimator.  However, they are both efficient.  The expected value of the standard deviation of the mean is 97\% of that $H\!L$ for normally distributed data.  The corresponding value for the median is 80\%.  The comparison for $Q_n$ and MAD are 88\% and 62\%.  We propose 
\begin{equation}
    R^2 = n \left [ H\!L \left ( X_i \right ) \right ]^2 + n \left [ Q_n \left (X_i \right ) \right ]^2
    \label{eq:chi2_robust}
\end{equation}
to fit the observed QPO frequencies. The $R^2$ estimator is an unbiased estimator for $\chi^2$ for normally distributed data.  Furthermore, the $H\!L$ and $Q_n$ estimators trade a modest amount of efficiency for robustness. They require slightly larger samples (about 40\% larger) to achieve the same precision on model parameters as the traditional $\chi^2$ estimator for normally distributed data, but are much less sensitive to outliers.

\begin{figure}
    \centering
    \includegraphics[width=\columnwidth]{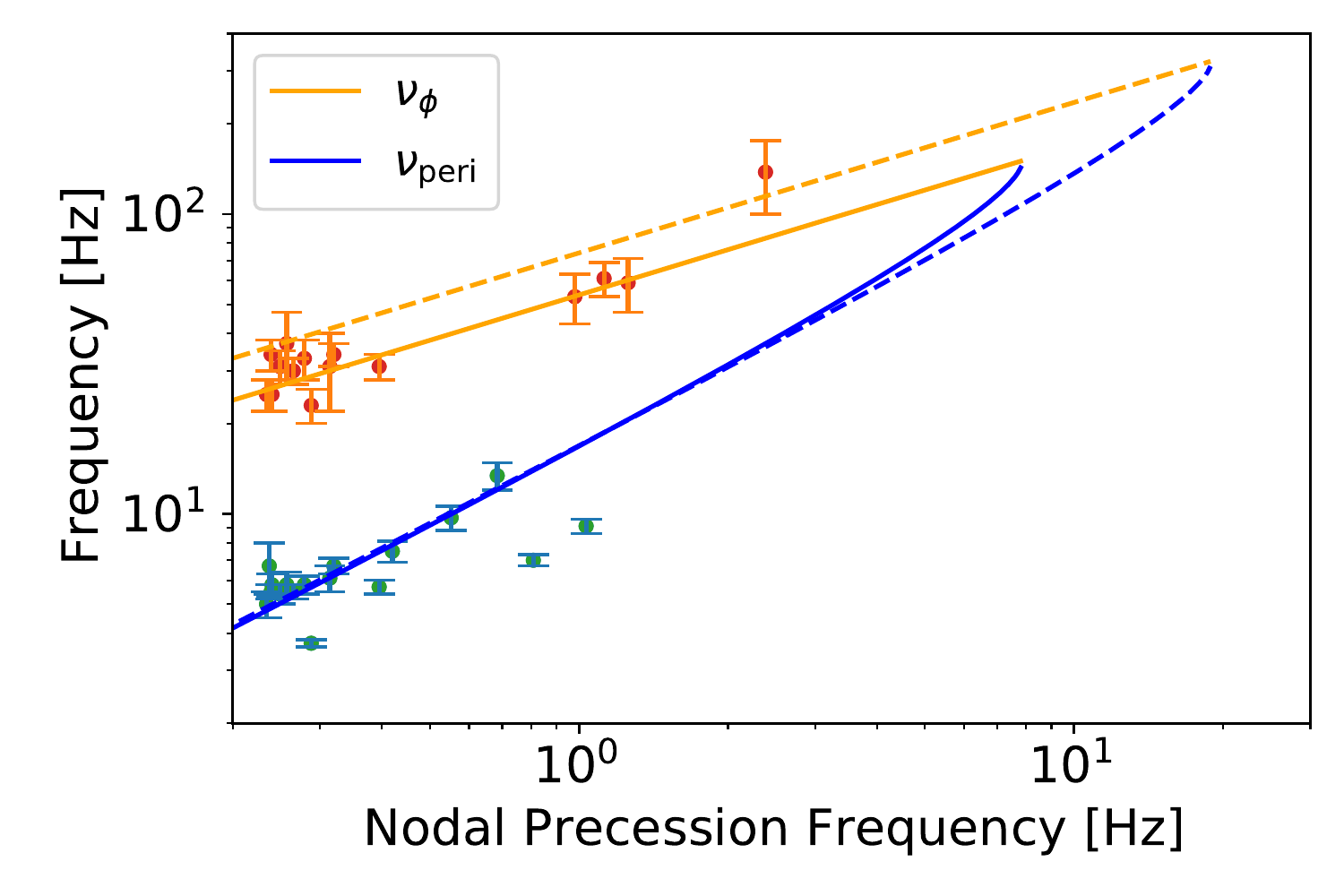}
    \caption{Same as Fig.\ref{fig:xte_chi2} but using the robust $R^2$ fitting technique. The solid curves trace the best-fitting model using the robust technique with
    $M=19.0~\textrm{M}_\odot$ and $a/M=0.31$. The value of $R^2$ is 57 compared to $\chi^2$ of 386 for the best fit in Fig.~\ref{fig:xte_chi2}.}
    \label{fig:xte_rob2}
\end{figure}

Fig.~\ref{fig:xte_rob2} shows the result of the fitting process.  The most obvious difference is in the model values for $\nu_{\textrm{\scriptsize peri}}$ that now avoid the two lower points that were apparent outliers in Fig.~\ref{fig:xte_chi2}.  This model has a lower spin of $a/M=0.31$, similar to the value of \citet{mottab} but a larger mass of $M=19.0 ~\textrm{M}_\odot$.  The value of $R^2$ is 57 compared to $\chi^2$ of 386 for the best fit in Fig.~\ref{fig:xte_chi2}.   Using the values derived by \citet{mottab} yields a value of $R^2$ of 185.   

Here we note a key qualitative difference between our fit and that of \citet{mottab}.  The value of the mass is significantly larger in our fit which means that the maximum value of the nodal precession frequency is less than 10~Hz.   \citet{mottab} presented measurements of single QPOs with frequencies up to nearly 20~Hz which they interpreted as the nodal precession frequency.  This interpretation requires a smaller mass than the one we obtained here.  The value of $R^2$ is similar to the number of degrees of freedom, 31.   This result is consistent with the fact that the bulk of the measurements are well characterized by the RPM and that the uncertainties have been well estimated.   We achieve this without explicitly identifying outliers.

\begin{figure}
    \centering
    \includegraphics[width=\columnwidth]{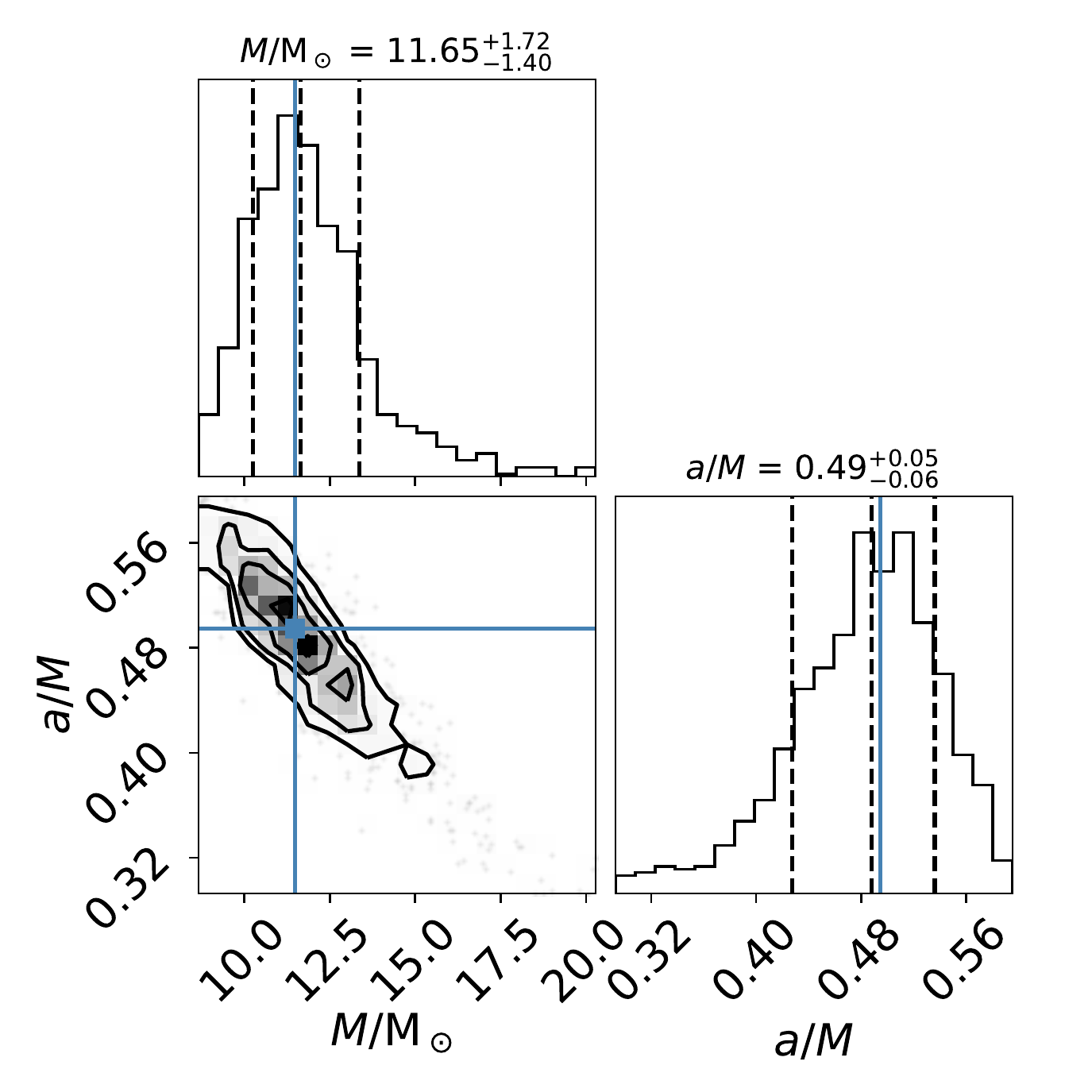}
    \caption{The results of the bootstrap resampling and $\chi^2$  analysis of the data for XTE~J1550-564.  The ranges depicted each contain 99\% of the values that results from the fits.  The blue lines indicate the best-fitting parameters for the original sample.}
    \label{fig:chi2_XTEcorner_GR}
\end{figure}

\begin{figure}
    \centering
    \includegraphics[width=\columnwidth]{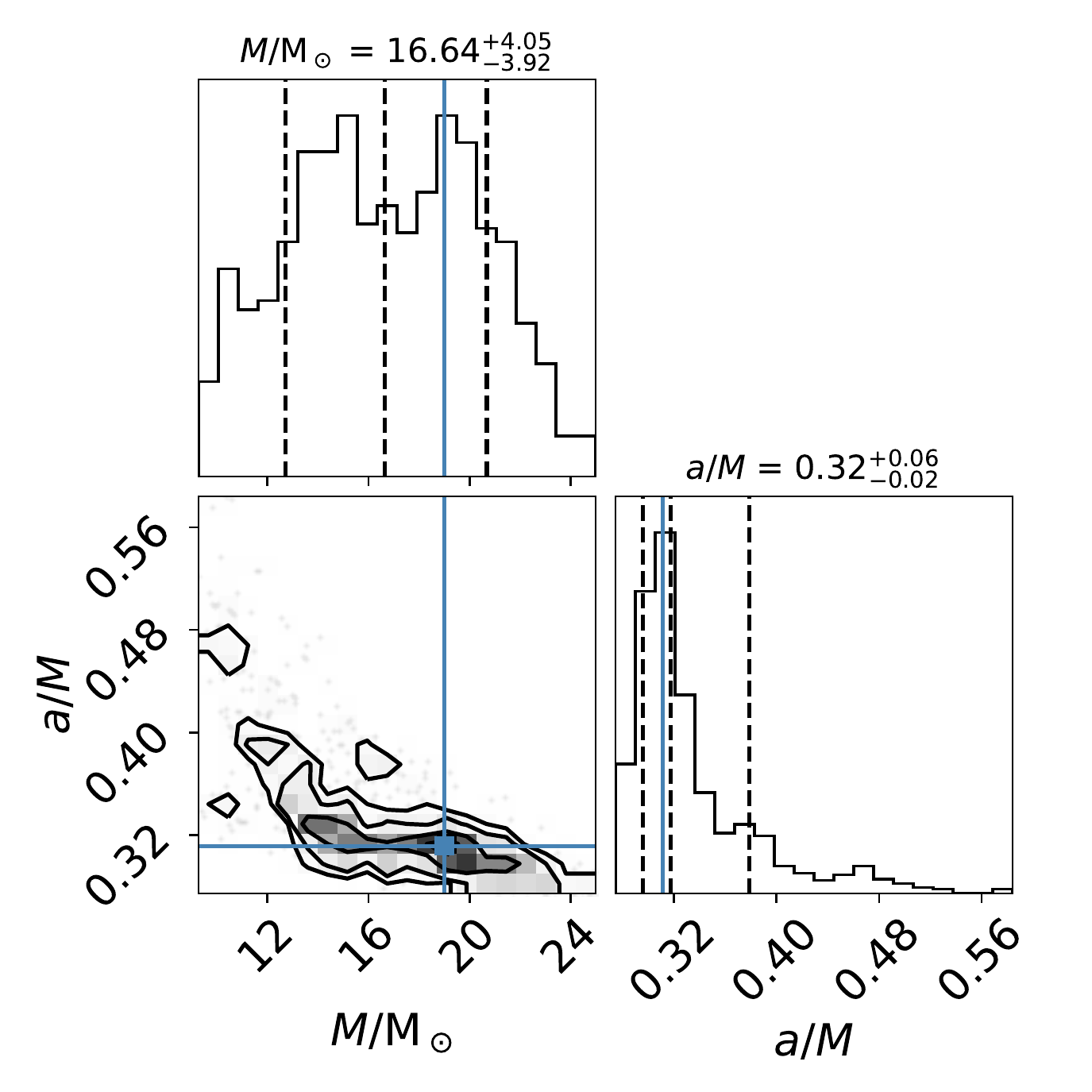}
    \caption{The results of the bootstrap resampling  and robust $R^2$ analysis of the data for XTE~J1550-564.  The ranges depicted each contain 99\% of the values that results from the fits.  The blue lines indicate the best-fitting parameters for the original sample.}
    \label{fig:robust_XTEcorner_GR}
\end{figure}

A crucial final part of the analysis is the estimation of the uncertainties in the results.  We estimate these uncertainties by a bootstrap resampling of the dataset, in which we select the same number of measurements as observed with replacement, so that the same measurement could be included many times or not at all.  Fig.~\ref{fig:chi2_XTEcorner_GR} and~\ref{fig:robust_XTEcorner_GR} depict results of the bootstrap resample for the $\chi^2$ and $R^2$ analysis.  The best-fitting values for the original dataset are depicted in blue for both techniques.  In particular, the $\chi^2$ analysis yields a smaller mass and larger spin than the $R^2$ analysis.  However, the best-fit $\chi^2$ model lies within the 99\% confidence region of the $R^2$ analysis and vice versa.  The uncertainties on the mass from the $R^2$ analysis are more than twice as large as for the $\chi^2$ analysis, but the uncertainties on the spin are comparable.   Although the constraints are weaker with the $R^2$ analysis, we choose to use it to constrain and quantify potential deviations from GR so that we can obtain the most conservative estimates of these quantities. 


\section{Results}

Our focus so far has been to demonstrate and explain our technique for analyzing the QPO data with XTE~J1550-564 as an example and by focusing on the standard GR parameters of mass and spin. However, in addition to XTE~J1550-564, \citet{mottaa} published an analysis of RXTE observations of GRO~J1655-40.  For this object, they found a simultaneous triplet of QPOs from which they could determine the mass and spin of the black hole (and the radius of the oscillations) from the RPM equations, yielding $M=5.3~\textrm{M}_\odot$ and $a/M=0.29$.  

\begin{figure}
    \centering
    \includegraphics[width=\columnwidth]{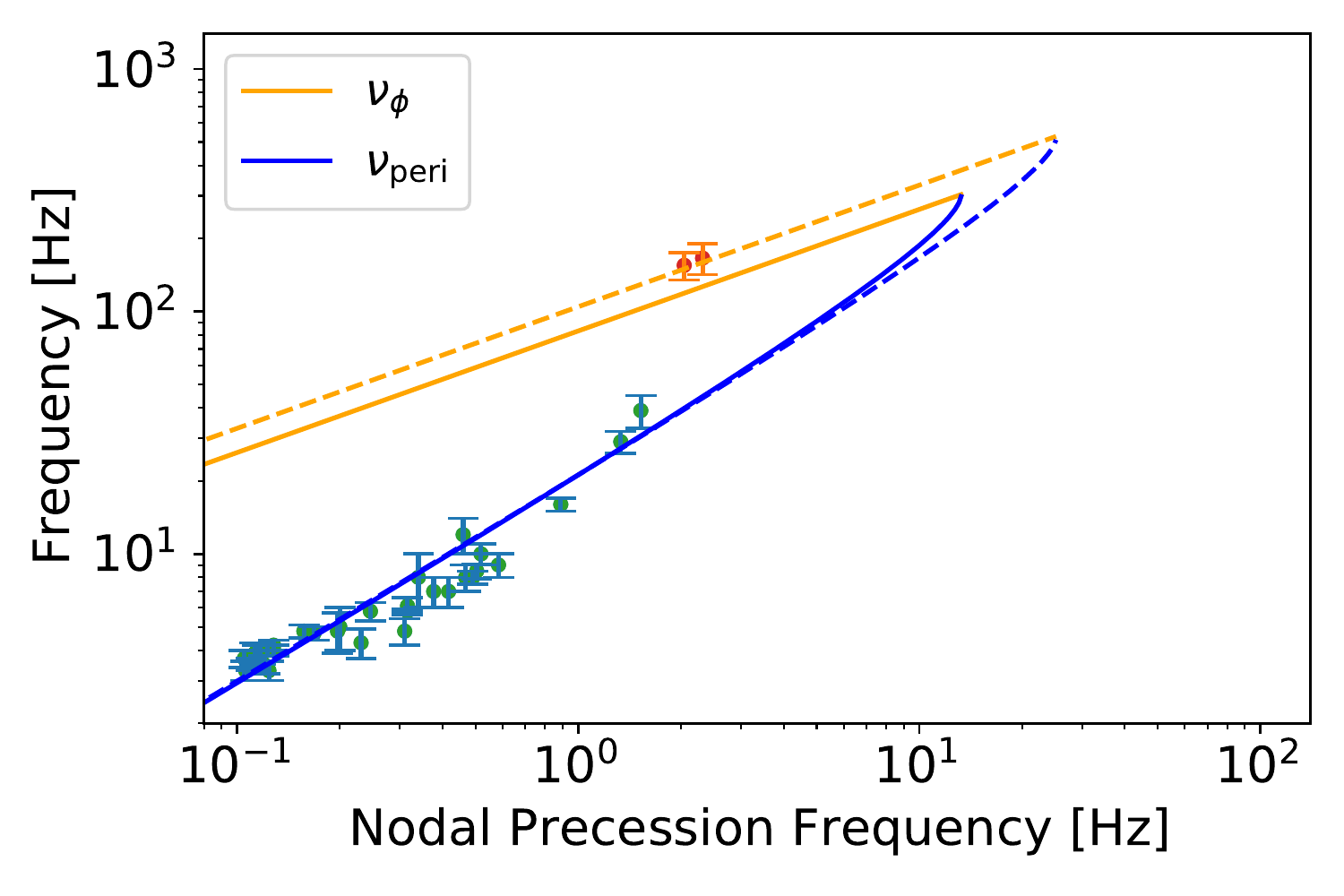}
    \caption{Same as Fig.\ref{fig:xte_chi2} but using the robust $R^2$ fitting technique for GRO~J1655-40. The dashed curves give the values predicted by the RPM for parameters determined by \citet{mottaa} $M=5.3~\textrm{M}_\odot$ and $a/M=0.29$. The solid curves trace the best-fitting model using the robust technique with $M=9.1~\textrm{M}_\odot$ and $a/M=0.27$.}
    \label{fig:gro_rob2}
\end{figure}

\begin{figure}
    \centering
    \includegraphics[width=\columnwidth]{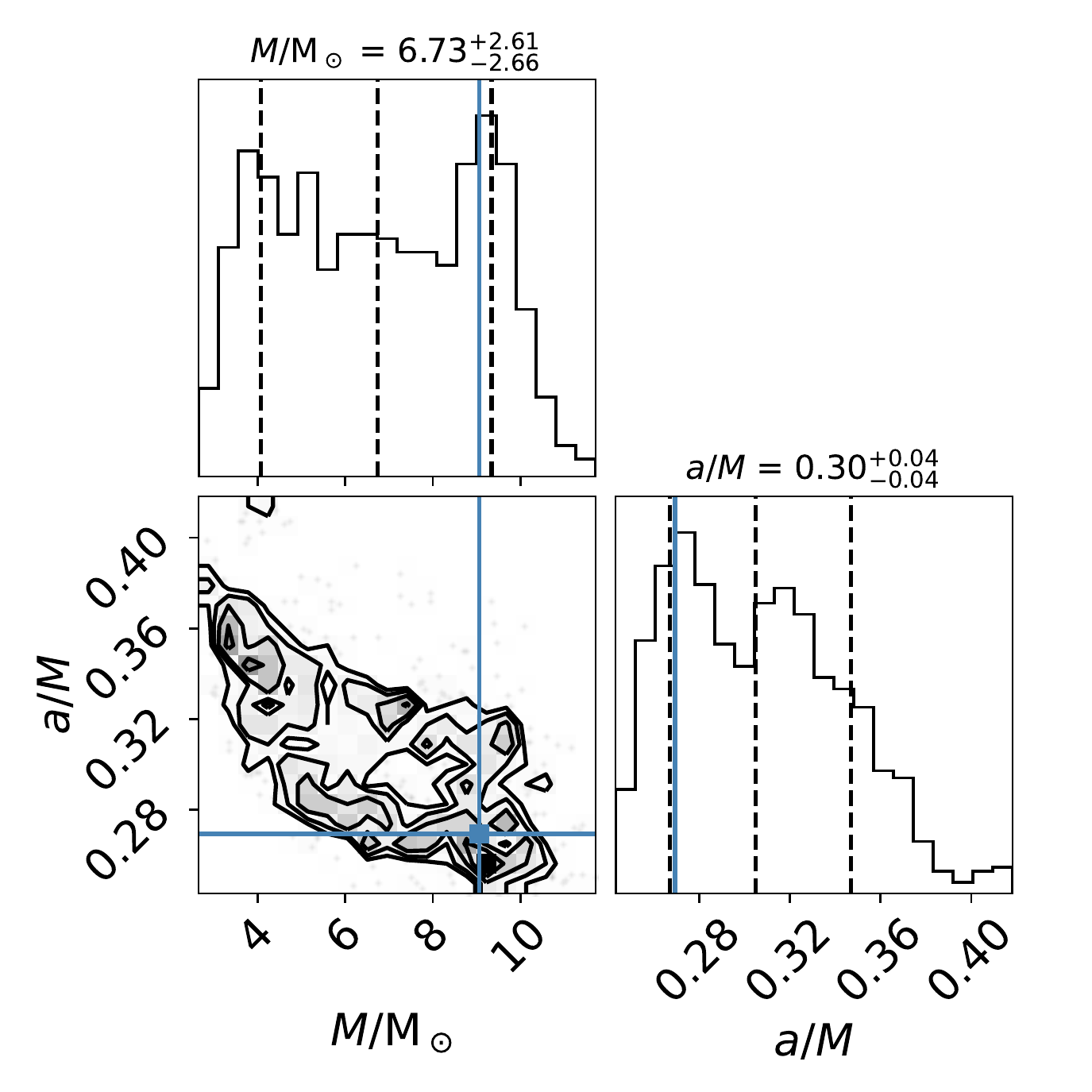}
    \caption{The results of the bootstrap resampling and robust $R^2$ analysis of the data for GRO~J1655-40.  The ranges depicted each contain 99\% of the values that results from the fits.  The blue lines indicate the best-fitting parameters for the original sample.}
    \label{fig:robust_GROcorner_GR}
\end{figure}

Fig.~\ref{fig:gro_rob2} depicts the best-fitting model for GRO~J1655-40 using the $R^2$ estimator compared with the model of \citet{mottaa}.  Again the robust $R^2$ estimator finds a larger mass resulting in smaller frequencies at the innermost stable orbit.  \citet{mottaa} present observations of single QPOs up to 25~Hz which would be difficult to reconcile with the best-fitting model if these single QPOs are manifestations of the nodal precession frequency.  Turning to the constraining power of the robust fitting technique, we find in Fig.~\ref{fig:robust_GROcorner_GR} that the distribution of bootstrapped samples is centered about $6.7~\textrm{M}_\odot$ and the values determined by \citet{mottaa} lie within the range of the fits.  

\begin{figure*}
    \centering
    \Large{GRO~J1655-40}

    \includegraphics[width=0.495\textwidth,trim=0 0 3.4in 0]{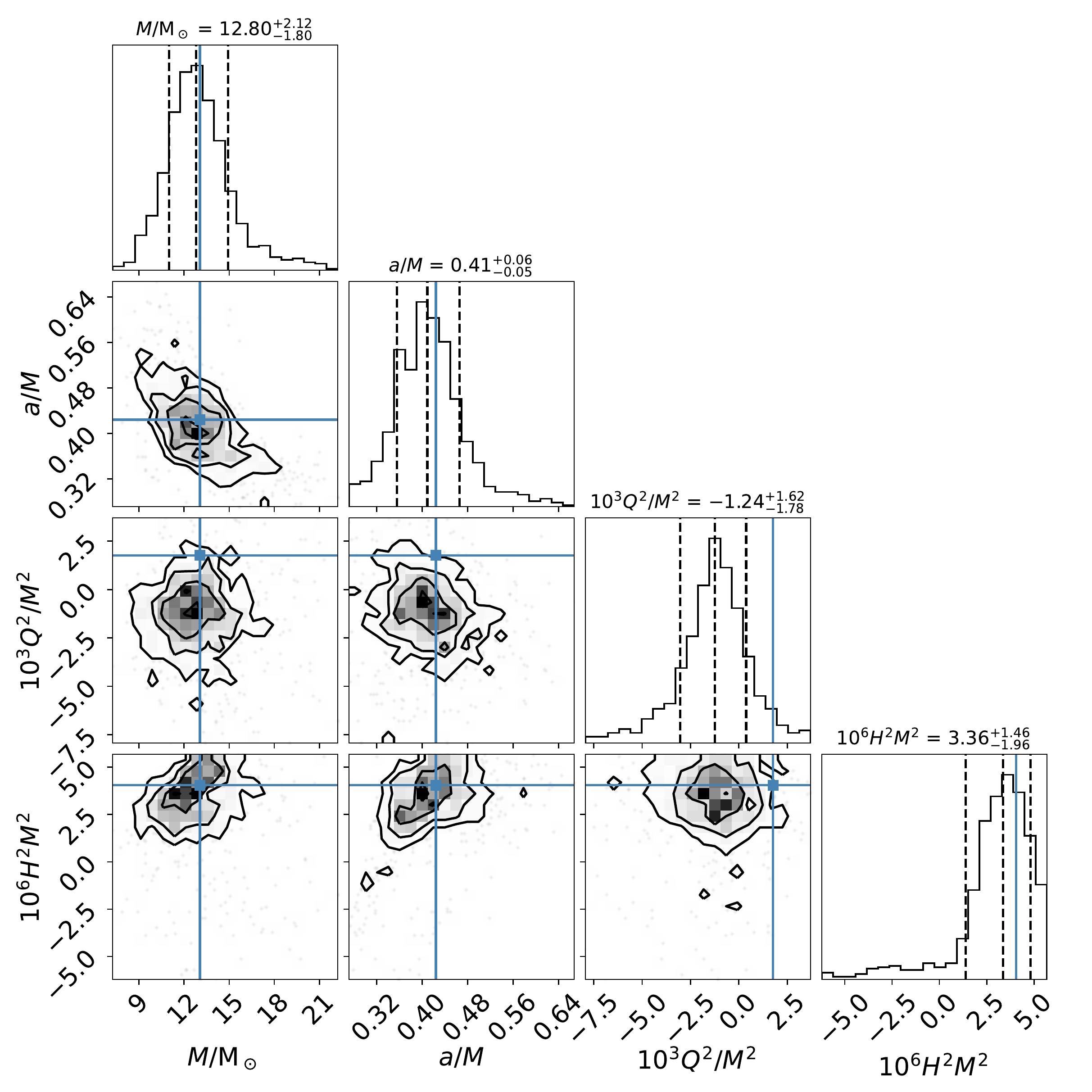}
    \includegraphics[width=0.495\textwidth,trim=3.4in -3in 0 0]{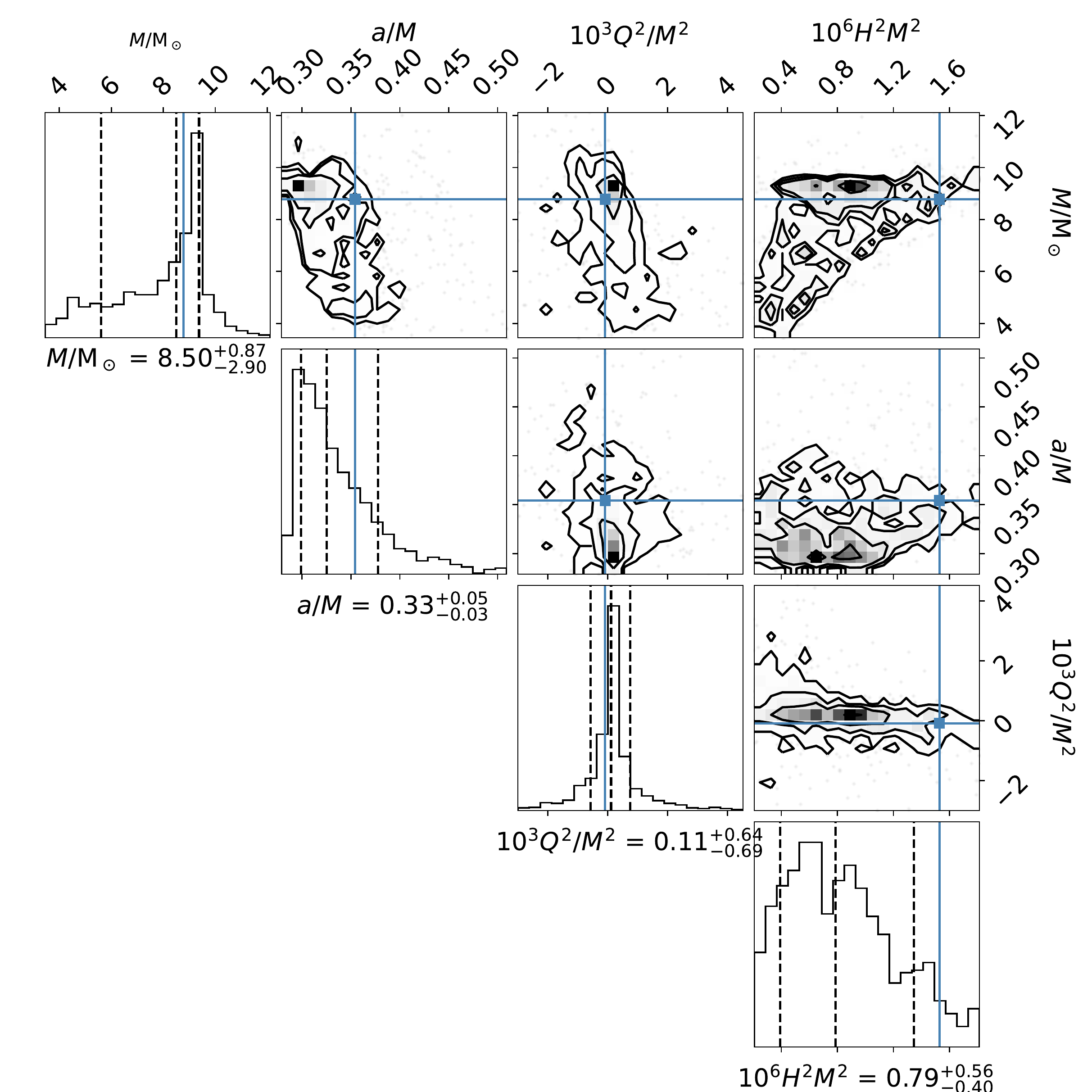}

    \Large{XTE~J1550-564}
    \caption{The results of the bootstrap resampling  and robust $R^2$ analysis of the data for XTE~J1550-564 (left) and GRO~J1655-40.  The ranges depicted for $M$ and $a/M$ each contain 99\% of the values that results from the fits.  The ranges for $Q^2/M^2$ and $H^2 M^2$ contain 95\% and 90\% of the values respectively. The blue lines indicate the best-fitting parameters for the original sample.}
    \label{fig:robust_Bothcorner_notGR}
\end{figure*}

We have been conservative in restricting the data to QPO doublets because the identification of the QPO frequencies as RPM frequencies is most certain in those cases and our ultimate goal is to derive constraints on GR, not the properties of the black hole for which the usefulness of the singlet QPOs trumps the uncertainties in their identification.  To place constraints on the additional metric parameters, we fit the mass, spin, charge and cosmological constant to the data and to the bootstrapped samples to obtain confidence regions on all of the parameters simultaneously, as depicted in Fig.~\ref{fig:robust_Bothcorner_notGR}.  In this broader class of models, the masses and spins are again typically larger than those obtained by \citet{mottaa} and \citet{mottab}, but the masses are smaller than obtained in the restricted GR fits.  

The key results here are the constraints on the parameters $Q^2$ and $H^2$ depicted in the figure in natural units of the black hole mass.  The value of $Q^2/M^2$ is constrained at the $10^{-3}$ level, and $H^2 M^2$ at the $10^{-6}$ level.  To understand the significance of these constraints on the components of the metric, we have to include the typical radius where the QPOs are excited within the model.  Using the observed Kepler frequencies and the best-fitting masses, we find that the QPOs are typically excited around a radius of $10 ~GM/c^2$. Comparing either the terms in the metric itself (Eq.~\ref{eq:spacetime}) or the square of the angular frequencies (Eq.~\ref{eq:Mfreq} to~\ref{eq:Hfreq}) yields the following fractional constraints on the deviation of the metric coefficients or frequencies from the GR values:
\begin{equation}
    \left | \frac{Q^2}{M r} \right | \lesssim 10^{-4}~\textrm{and}~  \left | \frac{H^2 r^3}{M} \right | \lesssim 10^{-3}
\end{equation}
where we have used $r=10 M$.  These constraints are the key result of this work.

\section{Summary and Conclusions}

Three types of QPOs have been observed coming from the accretion disks around black holes. Though the main mechanism behind these frequencies is still unknown, the Relativistic Precession Model \citep{RPM}  appears to match the data well when tested. This model was therefore used to test GR in the strong field regime using the Kerr-Newman-deSitter metric, which describes a spinning and charged black hole with the addition of a cosmological constant like parameter. 

Thus far, the cosmological constant has been assumed to be negligible close to the black-hole horizon.  However, if we are to find a theory that connects general relativity and quantum mechanics, it is essential that we understand each theory to their fullest extent and make no approximations in the coming age of enhanced instrumentation. We therefore used a nonzero cosmological constant parameter to constrain any type of excess energy density that could be in the vicinity of the black hole.

After calculating three frequency equations to describe the three observed QPOs in a Kerr-Newman-deSitter spacetime, our models were tested against RXTE data. We compared our fit using four black hole parameters to a fit using the standard Kerr metric as well as the fits derived in \citet{mottaa}.  Their fits relied on a single instance of simultaneously observed QPOs. This study fit for black hole parameters using models optimized to all observed QPOs, instead of using calculated values based on simultaneous QPO observations or values taken from the literature to test RPM.

Finally, statistical tests were conducted to compare the three models more precisely. We were able to make fractional constraints on deviations from GR at a radius of $10M$ to be  
$$
\left | \frac{Q^2}{M r} \right | \lesssim 10^{-4}~\textrm{and}~  \left | \frac{H^2 r^3}{M} \right | \lesssim 10^{-3}.
$$
Interpreting the results of our bootstrap resampling analysis, we see that the charge parameter heavily peaked around $Q^2 = 0$ for both sources, as expected (Fig.\ref{fig:robust_Bothcorner_notGR}). However, our cosmological constant parameter $H^2$ was rarely zero, which suggests that there could be excess energy density or some similarly behaving parameter yet unaccounted for in GR. It is also possible the nonzero nature of the $H^2$ parameter fit is a result of one or more QPOs being incorrectly classified.  

We conducted our analysis using robust methods for testing GR, prioritizing reduction in QPO identification uncertainty over the inclusion of more data and using fitting techniques insensitive to outliers. Our methods were therefore less reliable in recovering values for mass and spin that closely agreed with those derived from X-ray observations. This does not diminish the accuracy of our GR constraints as our best fit values fall within the estimated uncertainties of past studies that relied on a priori assumptions not considered here \citep{Orosz,beer}.

This study demonstrates that robust tests of GR are possible in the strong field regime using QPO observations. Our methods, when applied to higher resolution data, will help confirm the RPM theory and further constrain the spacetime around black hole regimes. More observations of coincident QPOs with precise timing would also help the accuracy of our $R^2$ analysis to further constrain our fits. 

Building on the growing literature of black hole X-ray timing analysis techniques, future studies will be able to compare observational results across models and further our understanding of the spacetime around black holes. These proof-of-concept studies are essential in attaining this goal if we wish to probe unknown parameter space and conduct precision tests of GR.




\section*{Acknowledgements}

This work has been supported by the Natural Sciences and Engineering Research Council of Canada through the Discovery Grants program and Compute Canada. \\

\begin{flushleft}
\textit{Data Availability} \medskip

The measurements used in this paper can be found in Tab. 1 of \citet{mottaa} and Tab. 3 of \citet{mottab}.

\end{flushleft}

\newpage

\bibliographystyle{mnras}

\bibliography{mainmn}

\label{lastpage}

\end{document}